# *Ackermann Encoding, Bisimulations, and OBDDs*

Carla Piazza

*Dipartimento di Informatica*
*Università Ca' Foscari di Venezia*
*Via Torino, 155 – 30172 Mestre (VE) – ITALY*
(*e-mail:* `piazza@dsi.unive.it`)

Alberto Policriti

*Dipartimento di Matematica e Informatica*
*Università di Udine*
*Via Le Scienze, 206 – 33100 UDINE – ITALY*
(*e-mail:* `policriti@dimi.uniud.it`)

## Abstract

We propose an alternative way to represent graphs via OBDDs based on the observation that a partition of the graph nodes allows sharing among the employed OBDDs. In the second part of the paper we present a method to compute at the same time the quotient w.r.t. the maximum bisimulation and the OBDD representation of a given graph. The proposed computation is based on an OBDD-rewriting of the notion of Ackermann encoding of hereditarily finite sets into natural numbers.

## 1 Introduction

In this paper we consider the problem of computing and representing the bisimulation on a given Kripke structure. Such a problem, central in Model Checking, has been tackled by many authors and various solutions have been given. In particular, the algorithm proposed by Kanellakis and Smolka (Kannellakis and Smolka 1990) is used in many model checkers with explicit-state representation (e.g., XEVE), while the algorithm proposed by Bouali and de Simone (Bouali and de Simone 1992) is used in the case of symbolic representation (e.g., NuSMV). The algorithms presented in (Bouajjani et al. 1990) and (Lee and Yannakakis 1992) are designed to obtain better performance in the case of the so-called On-the-Fly Model Checking. The routine proposed by Paige and Tarjan in (Paige and Tarjan 1987) is still the best in terms of worst case time complexity ($O(|E|\log|N|)$). In (Dovier et al. 2001) it has been proposed an algorithm which works on explicit representations and which, in the worst-case, has the same time complexity as the one by Paige and Tarjan, but which in many cases reaches a linear time complexity $O(|E|)$ and requires less space during the computation.

A particularly interesting line of research is witnessed by the work presented in (Bouali and de Simone 1992) focussed on the design of algorithms whose execution can be easily coordinated with efficient (symbolic) representation techniques for the underlying Kripke structures. The OBDD data structures (see (Bryant 1986)) employed in Symbolic Model Checking (see (McMillan 1993)), allow the storage in memory of much larger structures and set new standards for the size of Kripke structures to which Model Checking can be applied. However, it is not always immediate to map the further space-saving method of bisimulation reduction on OBDD-representation of Kripke structures: especially designed algorithms must be proposed for this purpose and different complexity parameters must be considered (Fisler and Vardi 1999; Somenzi 1999).

In this paper we propose an alternative way to represent graphs via OBDDs which is based on the use of a (generic) partition on the nodes and which allows sharing among the employed OBDDs. Instead of using an OBDD to represent the graph's relation, after splitting the graph into blocks, for a given block $B_j$ we use an OBDD whose size depends upon the list $D_j$ of nodes reachable in one step from $B_j$ only. If the graph has $2^u$ nodes, then the unique OBDD of the standard representation has $2u$ levels, while with our technique we use OBDDs with $u + k_j$ levels, where $k_j = \log |D_j|$. Moreover, some sharing becomes possible: for example if all the $D_j$s have approximately the same cardinality, then they share the last $k_j$ levels.

When the initial partition is based on a suitable notion of *rank* we are able to propose a method for bisimulation computation as well as OBDD-representation, i.e. the output of the routine is the bisimulation quotient represented using our OBDD-encoding. The technique is based on the fact that a Kripke structure is nothing but a (in general redundant) representation of an hereditarily finite set. To this end, starting from an encoding *à la* Ackermann (see (Ackermann 1937; Levy 1979)) of hereditarily finite sets as OBDDs, we propose an extension of such an encoding to non-well-founded hereditarily finite sets and a computation technique exploiting the *a priori* bound on the size of the set to be encoded for its determination. The key notion behind our encoding is an extension of the notion of *rank* to non-well-founded sets, already used in (Dovier et al. 2001) for explicit bisimulation computation.

The paper is organized as follows: in Section 2 we present the problem; in Section 3 we review some related works; in Section 4 we consider techniques for OBDDs representation of graphs; in Section 5 we propose the alternative representation, based on a layering of the graph, for which the basic operations are discussed in Section 6; in Section 7 we propose the method to compute the OBDD representation of the quotient of an acyclic graph w.r.t. the maximum bisimulation; in Section 8 we complete our discussion extending the method to the cyclic case. Some concluding remarks, including a short discussion on the similarity between Ackermann encoding and OBDD representation, end the paper.

# PART ONE
# Basics

## 2 Preliminaries

*Definition 1* (*Rooted Graph*)
A (direct) *rooted graph*[1] is a triple $G = \langle N, E, r \rangle$, where $N$ is the set of nodes, $E \subseteq N \times N$ is the set of edges, and $r \in N$ is a node such that all the other nodes in $N$ are reachable from $r$, i.e. for all $a \in N$ there is a sequence $a_1, \ldots, a_h$ of elements of $N$ such that $rEa_1$, $a_h = a$, and $a_i E a_{i+1}$ for all $1 \leq i \leq h-1$.

From now on and for reasons that will become clear below, the relation $E$ will be denoted by $\succ$ (a stylized format for both $\ni$ and $\rightarrow$) and its inverse relation $E^{-1}$ will be denoted by $\prec$. We will use the term graph to refer also to rooted graphs.

In the rest of this paper we will mostly deal with the following problem:

> *Given a graph $G = \langle N, \succ, r \rangle$ determine a 'compact' representation of $G/\equiv$, where $\equiv$ is the maximum bisimulation over $G$.*

The problem is well-known in the area of Model Checking, where graphs have usually labels on nodes. We briefly recall the definition of bisimulation on graphs with and without labels and we justify the fact that in this paper we deal with graphs *without* labels.

*Definition 2* (*Labelled Graph*)
Let $\Sigma$ be a finite alphabet. A *labeled graph* is a triple $G = \langle N, \succ, r, \ell \rangle$, where $\langle N, \succ, r \rangle$ is a rooted graph and $\ell : N \rightarrow \Sigma$ is a labeling function.

*Definition 3* (*Bisimulation*)
Given a labeled graph $G = \langle N, \succ, r, \ell \rangle$ a *bisimulation* is a relation $B \subseteq N \times N$ which satisfies:

(label)  $a \, B \, a' \Rightarrow \ell(a) = \ell(a')$;
(forw.)  $a \, B \, a' \wedge a \succ b \Rightarrow \exists b'(a' \succ b' \wedge b \, B \, b')$;
(back.)  $a \, B \, a' \wedge a' \succ b' \Rightarrow \exists b(a \succ b \wedge b \, B \, b')$.

Given a graph $G = \langle N, \succ, r \rangle$ a bisimulation is a relation $B \subseteq N \times N$ which satisfies the conditions (forward) and (backward).

We have given the definition of bisimulation $B \subseteq N \times N$ on a graph $G = \langle N, \succ, r \rangle$, but it is immediate to imagine the definition of a bisimulation $B \subseteq N_1 \times N_2$ between two graphs $G_1 = \langle N_1, \succ_1, r_1 \rangle$ and $G_2 = \langle N_2, \succ_2, r_2 \rangle$: it is, essentially, only necessary to add the condition that $r_1 B r_2$.

The main theorem on maximal bisimulations states that:

*Theorem 4*
Given a graph $G$ (with or without labels) there always exists a (unique) maximum bisimulation $\equiv$ which is an equivalence relation.

---

[1] In this paper we always refer to finite graphs.

*Proof*
See (Aczel. 1988). □

We will say that two nodes $a, a'$ of a graph are *bisimilar* if and only if $a \equiv a'$.

The quotient structure $G/\equiv$ can be seen as the (most compact) graph representation of the non-well-founded set (see (Aczel. 1988)) associated to the root of $G$ and with $\succ$ acting as the membership relation $\ni$, (Dovier et al. 2001).

Given a graph $G' = \langle N', \succ', r', \ell \rangle$ it is possible to encode the labeling on the nodes by adding new nodes to the graph and obtaining a new graph $G = \langle N, \succ, r \rangle$ in such a way that there is a complete correspondence between the maximum bisimulation over $G'$ and the maximum bisimulation over $G$. In some applications the graphs also have labels on the edges and the definition of bisimulation takes into consideration both the labels on the edges as well as the labels on the nodes. By adding new nodes it is possible to encode also the labeling of the edges. A possible (linear) encoding for the labels on the edges and on the nodes is described in (Dovier et al. 2001): hence, it will not be restrictive to consider the case of graphs without labels.

In this paper we propose a method to determine an OBDD representation of the quotient under the maximum bisimulation of a given graph $G = \langle N, \succ, r \rangle$. If one prefers not to move from a graph $G$ to its unlabelled $G'$ it is straightforward to extend the technique in order to cope directly with the labels. In the following we first give some references about related work, then we start with some reflections about symbolic representations and we propose an OBDD representation based on a partitioning of $N$ which is at the core of our method to compute the maximum bisimulation.

## 3 Related work

The following material is related with both OBDDs and bisimulation.

OBDDs, *ordered binary decision diagrams*, are a canonical representation for boolean functions, i.e. two boolean functions are equivalent if and only if they are associated to the same OBDD. General BDDs were first introduced in (Lee 1959) and in (Akers 1978). Bryant, defining in (Bryant 1985) the more restricted notion of OBDDs attracted attention to their use in *logic design verification*.

OBDDs are used in Model Checking to represent the labelled graph which models the behavior of the system. Such a representation, usually called *symbolic* or implicit representation, allows to deal with systems with much more states than an explicit representation as noticed in (Burch et al. 1992).

Unfortunately, it is possible, in the worst case, that the symbolic representation of a system is as large as the explicit one, see (Somenzi 1999). Many authors have tried to solve this problem using different techniques. The size of an OBDD depends on the ordering of the variables, hence methods to determine "a good" *variable ordering*, based on the use of heuristics which try to exploit the structure of the system representation (Malik et al. 1988) and on dynamic reordering (Rudell 1993), have been proposed. However, there are many applications where these optimization techniques for OBDDs reach their limits. For a fixed state encoding there are many finite state machines whose OBDD representations are large regardless of the variable orders, see (Aziz et al. 1994). When OBDDs are used to represent graphs the size of the OBDD depends also on the state encoding, and to this end in (Meinel and Theobald 2001) *local* encoding transformations have been studied.

Another well-known technique to reduce the size of the OBDDs consists in *partitioning* the OBDD (Burch et al. 1991; Meinel and Stangier 2000; Meinel and Stangier 2001), in the sense that, instead of using a unique OBDD to represent the transition relation of a graph, one OBDD for each variable of the target nodes is considered: the global OBDD is obtained from all these OBDDs. Such partitioning is substantially different from the partitioning we propose here: all the source nodes occur in all the OBDDs, while in our approach we partition the source nodes and each of them occurs only once.

Further attempts to reduce the complexity of the OBDD-representation can be found in (McMillan 1996; Cabodi 2001) where various form of decompositions based on the use of complicated functions to combine OBDDs are defined.

As far as bisimulation is concerned it is difficult to accurately list all the fields in which, in one form or another, the notion of *bisimulation* was introduced and now plays a central role. Among the most important ones are: Modal Logic, Concurrency Theory, Set Theory, and Formal Verification. In Model Checking several existing verification tools make use of bisimulation in order to minimize the state spaces of systems description and to check equivalence between transition systems. The verification environment XEVE (Bouali 1998) provides bisimulation tools which can be used for both minimization and equivalence test. In general, in the case of explicit-state representation, the underlying algorithm used are the ones proposed by Kanellakis and Smolka (Kannellakis and Smolka 1990) and by Paige and Tarjan (Paige and Tarjan 1987), while Bouali and de Simone algorithm (Bouali and de Simone 1992) is used in the case of symbolic representation. All these algorithms are based on *negative strategies*: start by considering that all the nodes bisimilar and separate the nodes when it is possible to prove that they are not bisimilar. On the contrary, a *positive strategy* would start by considering that all the nodes not bisimilar and put together nodes when they have been proved to be bisimilar. In (Dovier et al. 2001) an algorithm for explicit representation which combines positive and negative strategies exploiting the notion of rank as been proposed. A symbolic version of the result in (Dovier et al. 2001) has been proposed in (Dovier et al. 2002) and led to the development of a linear symbolic algorithm for strongly connected components computation (see (Gentilini et al. 2003)). The method we present here differs from all the previous ones because it is based on a fully positive strategy and it strongly exploits the notion of rank, together with the alternative symbolic representation we introduce in order to perform the bisimulation computation.

## 4 OBDDs representing graphs

The way OBDDs are usually employed in Model Checking to represent the states' space $N$, sets of states $S \subseteq N$, and the transition relation $\gg$, is based on the following observations (Clarke et al. 1999):

- we can safely assume that $N = \{0,1\}^u$, i.e. each node is encoded as a binary number;
- a set $S \subseteq N$ is a set of binary strings of length $u$, hence its characteristic function $\chi_S : \{0,1\}^u \to \{0,1\}$, where

$$\chi_S(s_1, \ldots, s_u) = 1 \quad \Leftrightarrow \quad \langle s_1, \ldots, s_u \rangle \in S$$

  is a boolean function, which can be represented using an OBDD;
- $\gg \; \subseteq N \times N$ is a set of binary strings of length $2u$ and hence, again, its characteristic function

$$\chi_\gg(x_1, \ldots, x_u, y_1, \ldots, y_u) = 1 \quad \Leftrightarrow \quad \langle x_1, \ldots, x_u \rangle \gg \langle y_1, \ldots, y_u \rangle$$

  is a binary function, which can be represented using an OBDD.

In particular, in the OBDD representing $\gg$ (without variable reordering) the first $u$ levels (variables) represent the codes of the source nodes, while the second $u$ levels (variables) represent the codes of the target nodes (see Figure 1).

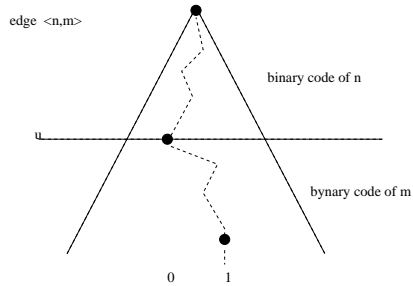

Fig. 1

*Example 5*
Consider the graph $G$ in Figure 2 on the left. Using the binary variables $x_1$ and $x_2$ for the code of the first node and the variables $y_1$ and $y_2$ for the code of the second node we obtain that the characteristic function of this graph is

$$(\neg x_1 \land x_2 \land \neg y_1 \land \neg y_2) \lor (x_1 \land x_2 \land \neg y_1 \land y_2) \lor$$
$$(x_1 \land \neg x_2 \land \neg y_1 \land \neg y_2) \lor (x_1 \land \neg x_2 \land \neg y_1 \land y_2)$$

which is represented by the OBDD in Figure 2 on the right.

If we use this OBDD encoding to represent $G/\equiv$ (without using variable reordering or other minimization techniques) we always obtain that if $|N/\equiv| = 2^{v-1}$, then the higher half of the OBDD is a complete tree with $2^{v-1} - 1$ nodes at the level corresponding to $y_1$ (the first variable relative to the encoding of the target nodes, see Example 5). This is because two paths of length $v$ which converge would represent two or more different nodes whose sets of $\succ$-successors are the same. However, since different nodes in $G/\equiv$ are not bisimilar, different nodes cannot have the same set of $\succ$-successors.[2] Even in case $G$ is not quotiented w.r.t. bisimulation, we obtain that (again, without using variable reordering) in the OBDD representing $\succ$, at level $v$ there are at least $|N/\equiv|-1$ different nodes.

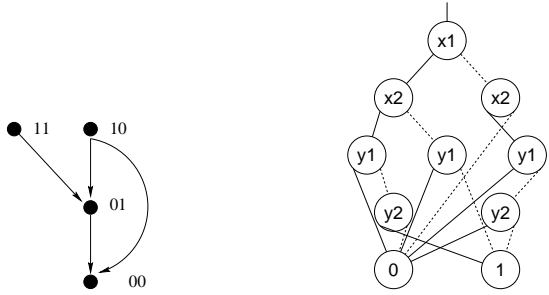

Fig. 2

A second disadvantage of this (rather classical) way of using OBDDs to represent a graph, is given by the fact that "topological" repetitions in the structure of the graph $G$ are not exploited in order to obtain further reductions. OBDDs are designed to reduce the dimensions of a graph "horizontally": nodes of the OBDD at the same level are collapsed only when the sub-OBDDs rooted on them are equal. In general there is no way to take advantage of a situation as the one described in the following example.

*Example 6*
*Consider the graph in Figure 3 on the left. The situation in the part A and B of*

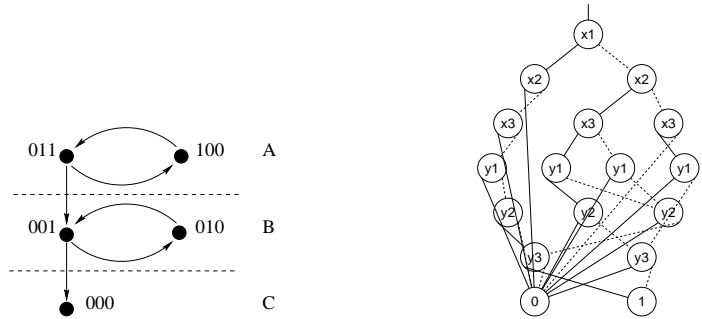

Fig. 3

*the graph is similar, but this has no reduction effects on the corresponding OBDD (see Figure 3 on the right) which at level* 3 *has* $|N|-1$ *nodes.*

---

[2] If there is a node without outgoing edges, then its path reaches directly 0, this is the reason for which we have $2^{v-1}-1$ instead of $2^{v-1}$ nodes at level $v$.

A "vertical" reduction could be performed only if it were possible to "ignore" some level, which means *do not care about the values of the variables on the levels that are ignored* (see below).

What we try to do in the next section is to propose an alternative way to use OBDDs in order to represent graphs in such a way to obtain:

- a vertical reduction, i.e. we deal with OBDDs shorter than $2u$, where $u = \log |N|$;
- use topological similarities to obtain further horizontal reductions.

In general, however, the global height of the OBDD remains $2u$, which implies that, in the worst case, it is possible to have $O(2^{2u})$ internal nodes in it.

## 5 OBDDs representing graphs: layering and sharing

As we observed, if we work on a graph $G$ reduced by bisimulation, we can safely assume that no two different nodes have the same set of successors with respect to $\succ$. Hence, the graph can be represented as a *collection of sets of $\succ$-successors* (one for each node). Notice that the so-called *unique-table* of most OBDD packages already performs this kind of optimization, see (Somenzi 1999).

Moreover, if we have a (generic) partition $P$ on the graph, each *block $B_j$* (*P*-class) of the partition can be represented separately, allowing sharing among the OBDDs used in the representation.

Below we will combine the above two ideas into an OBDDs representation minimizing the size of the data structures involved. The technique will turn out to be suitable for use even *while* the quotient structure $G/\equiv$ is determined starting from $G$ (cf. Sections 7 and 8).

The following definitions will be used:

- let $P = \{B_1, \ldots, B_p\}$ be a partition of $N$;
- for each block $B_j \in P$ consider the list $D_j$ of the nodes reached from a node in $B_j$:

$$D_j = [b \mid \exists a \in B_j (a \succ b)],$$

  Let us assume the nodes in $D_j$ are ordered using their binary codes and let $|D_j| = h_j$;
- for each $b \in D_j$ let $d_j(b)$ be the binary representation of the position in which $b$ occurs in $D_j$ ($d_j(b)$ is in $\{0,1\}^{\lfloor \log h_j \rfloor + 1}$);
- for each $a \in B_j$ consider the boolean function $\succ_j(a) : \{0,1\}^{\lfloor \log h_j \rfloor + 1} \to \{0,1\}$ defined as

$$\succ_j(a)(z_1, \ldots, z_{\lfloor \log h_j \rfloor + 1}) = \bigvee_{a \succ b} \chi_{\{d_j(b)\}}(z_1, \ldots, z_{\lfloor \log h_j \rfloor + 1}),$$

  where $\chi_X$ is the characteristic function of $X$.

*Remark 7*
It is clear that if we know $P$, all the $D_j$'s, and all the $\succ_j(a)$, we can infer $\succ$.

Moreover, the reader can check that if $P = \{N\}$, the representation proposed collapses to the technique discussed in the previous section. For each node $a$ we give the OBDD rooted at the end of the binary code of $a$. In the case $P = \{N\}$ the connection between the representation presented in the previous section and the representation we propose is similar to the connection between the adjacency-matrix and the adjacency-list representations of a graph.

The data structures that can be used to keep all these information are the following:

- the binary encodings of the nodes in $N$, where we assume that the first $q = \lfloor \log p \rfloor + 1$ ($p$ is the number of blocks) digits represent the block $B_j$ to which $a$ belongs;
- the following set of pairs:

$$\mathcal{D}_j = \{\langle b, d_j(b) \rangle \mid b \in D_j\},$$

  is a function associating to each element in $D_j$ its position, whose characteristic function $\chi_{\mathcal{D}_j}$ can be represented using an OBDD;
- the boolean function $\succ_j(a)$ can be represented using an OBDD (to which $a$ points).

All the OBDDs can be kept in a unique table with no duplicate sub-OBDDs, i.e. it is possible that the OBDD of $\mathcal{D}_j$ and the OBDD of $\mathcal{D}_{j'}$ share some sub-OBDDs. In particular, whenever $\mathcal{D}_j$ and $\mathcal{D}_{j'}$ have the same cardinality they share all the levels relative to the $d_j(b)$'s.

Using the above representation with a non-trivial partition $P$: the OBDDs of the functions $\succ_j(a)$'s are shorter than $u$ (vertical reduction); it is possible that $a$ is not bisimilar to $a'$, but $\succ_j(a) = \succ_{j'}(a')$, i.e. they point to the same OBDD (reduction due to topological repetitions).

It immediately follows that the set of nodes which refer to the same OBDD can be itself represented using an OBDD of height $u$.

*Example 8*
Consider the graph in Figure 3 (see Example 6). Using the partition $P = \{A, B, C\}$ we obtain.

$$D_C = \emptyset \quad D_B = [000, 001, 010] \quad D_A = [001, 011, 100]$$

$$\succ_C(000) = 0$$
$$\succ_B(001) = (\neg z_1 \wedge \neg z_2) \vee (z_1 \wedge \neg z_2) \quad \succ_B(010) = (\neg z_1 \wedge z_2)$$
$$\succ_A(011) = (\neg z_1 \wedge \neg z_2) \vee (z_1 \wedge \neg z_2) \quad \succ_A(100) = (\neg z_1 \wedge z_2)$$

Hence 001 and 011 share the same OBDD (and similarly 010 and 100), in particular 3 nodes are sufficient. In Figure 4 we show the OBDDs relative to $\mathcal{D}_A$ and $\mathcal{D}_B$ which can share entirely the last two levels.

Notice we can still use all the optimization techniques that are provided to deal

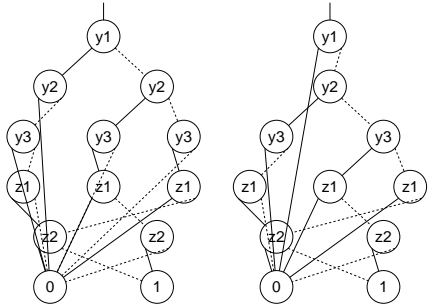

Fig. 4

with OBDDs and, moreover, we are not forced to keep all the OBDDs in central memory.

If $|N| = 2^u$, *out* is the maximum number of outgoing edges from a node, and we partition $N$ in $2^q$ classes with $q \sim u$ (using the first $q$ digits), then we obtain that:

1. in each class there are $2^{u-q} = k$ elements, with $k << |N|$;
2. each $D_j$ has at most $out * k$ elements;
3. the OBDDs of the functions $\succ_j(a)$'s have at most $\lfloor \log(out * k) \rfloor + 1$ levels.

If $out << |N|$, then we obtain that $out * k$ can be considered as a constant w.r.t. $|N| = 2^u$, the OBDDs of $\mathcal{D}_j$'s have about $u$ levels[3], and a large number of OBDDs of $\succ_j(a)$'s are shared.

The representation achieves good results, in particular, if there are few outgoing edges from each node or if there are topological repetitions. This happens, for instance, in the case of graphs obtained from programs to be analyzed when: a procedure is used more than once; different procedures perform symmetric actions; there are sequences of deterministic iterations.

*Example 9*
*Consider a graph with $2^u$ nodes composed by $2^q$ cliques of $2^{u-q}$ nodes with $u - q << u$. Moreover the cliques are connected in a cycle as shown in Figure 5 in the case $u = 5$ and $q = 3$. This graph can be interpreted as the representation of a communication process in which all the agents belonging to the same clique are able to communicate and only two special agent in each clique are able to communicate directly with another clique. In this case, with the standard representation is necessary to use one OBDD with $2u$ levels. Using our representation it is sufficient to use: $q$ OBDDs with $u$ levels and one shared OBDD with $u - q + 1$ levels to represent the $D_i$'s; 3 OBDDs with $u - q + 1$ levels to represent the $\succ_j(a)$'s; 3 OBDDs with $u$ levels to represent which nodes share the same OBDD. Notice that the constant 3 is due to the fact that in each clique $C_j$ there are three kind of nodes: the nodes which reach only the nodes in the clique, the node which reaches the clique $C_{j-1}$, and the node which reaches the clique $C_{j+1}$. Hence, using the standard representation, depending on the encoding of the nodes and on the variable ordering, it is possible that there*

---

[3] Exactly $u + \lfloor \log(out * k) \rfloor + 1$ levels.

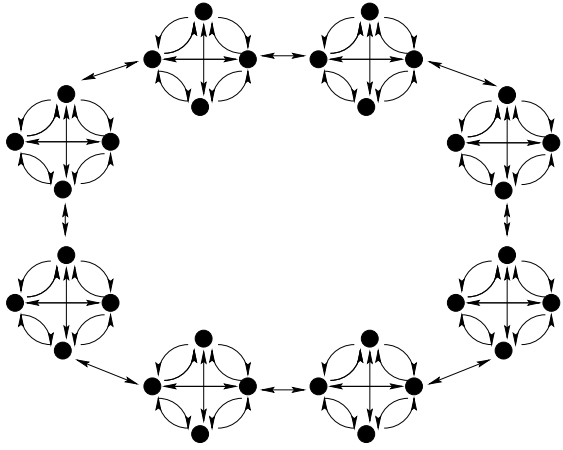

Fig. 5

are $2^{2u}$ internal nodes in the OBDD, while using our representation there are at most $2^{\log q + u - q + 1} + 3 * 2^{u-q+1} + 3 * 2^u \leq 2^{u+3}$ internal nodes.

Another example in which it is immediate to verify that our representation in the worst case works better than the classical one, is the case of graphs that are downward $n$-ary trees. In this case, partitioning the nodes into blocks of cardinality $k * n$ the $D_j$'s have cardinality $k$.

## 6 Operations

The operations on the representation we propose are similar to the ones on the classical representation with OBDDs. In this section we briefly discuss how to compute the image and the counter-image of a set of states.

Let $S \subseteq N$ be a set of states. Assume that we want to compute $\succcurlyeq(S) = \{b \,|\, \exists a \in S(a \succcurlyeq b)\}$ (image computation). We have that $S = S_1 \cup \ldots \cup S_p$, where each $S_i$ is a subset of a block in $P$ and we obtain that

$$\chi_{\succcurlyeq(S)}(\bar{y}) \;=\; \bigvee_{i=1}^{p} \exists \bar{z} [(\bigvee_{a \in S_i} \succcurlyeq_i (a)(\bar{z})) \wedge \chi_{\mathcal{D}_i}(\bar{y}, \bar{z})].$$

Hence, in order to iteratively apply this operation (reachability computation) it is necessary to intersect $\chi_{\succcurlyeq(S)}(\bar{y})$ with all the blocks of $P$. In general, given a set $S$ of states its intersection $S_i$ with the block $B_i$ of $P$ is represented by the OBDD of the function

$$\chi_{S_i}(\bar{x}) \;=\; \chi_S(\bar{x}) \wedge \chi_{B_i}(\bar{x}).$$

If we assume that $P$'s blocks are defined using the first $q$ digits of the codes of the nodes, then the elements of $S_i$ are the nodes in $S$ whose code start with the binary representation of $i$.

In order to compute $\prec(S) = \{b \mid \exists a \in S (b \succ a)\}$ (pre-image computation), we have to compute

$$\begin{aligned} g_B(\bar{z}) &= \exists \bar{y}(\chi_S(\bar{y}) \wedge \chi_{\mathcal{D}_B}(\bar{y}, \bar{z})) \\ \chi_{\prec(S)}(\bar{x}) &= \bigvee_{B \in P} \bigvee_{\bar{x} \in B} (\exists \bar{z}(g_B(\bar{z}) \wedge \succ_B(\bar{x})(\bar{z}))) \end{aligned}$$

Using similar boolean functions it is possible to move from the representation we propose to the classical one and vice-versa.

In the next section we concentrate on a computation of the previously proposed representation for a symbolic version of $G/\equiv$, obtained starting from a representation of $G$.

<div align="center">PART TWO</div>

# An OBDD representation of $G/\equiv$

### 7 Acyclic case

In order to obtain the OBDDs representation *while* computing $G/\equiv$, we will use an idea that can be traced back to the so-called *Ackermann encoding* of hereditarily finite sets into natural numbers (Ackermann 1937; Levy 1979). Such an encoding, inductively defined as

$$A(a) = \Sigma_{b \in a} 2^{A(b)},$$

establishes a bijection between the collections of hereditarily finite sets and the natural numbers. Hence, two sets (possibly represented by different means) will be bisimilar if and only if they are mapped by $A$ into the same natural number. Therefore, the computation of $A$ and the bisimulation relation are naturally carried on *together*. In our approach we essentially redefine $A$ using OBDDs in place of natural numbers and layering the definition on the ranks in order to take advantage of repetitions as illustrated in Section 5. Given a node $a$ we will call this OBDD rank-based Ackermann encoding $\mathbb{A}(a)$

We start from the acyclic case because in this case the notion of rank partitions the graph $G$ and gives an order between the classes of such a partition which allows to compute the OBDD representation of $G/\equiv$.

*Definition 10* (*Rank - acyclic case*)
Given an acyclic graph $G = \langle N, \succ, r \rangle$, the *rank* of a node $a \in N$ is defined as:

$$rank(a) = \begin{cases} 0 & \text{if } \forall b \in N \neg(a \succ b) \\ max\{rank(b) \mid a \succ b\} + 1 & \text{otherwise} \end{cases}$$

The following lemma states the main property (for our application) of the rank.

*Lemma 11*
Let $G$ be an acyclic graph and $a, a' \in N$ be two nodes of $G$.

$$a \equiv a' \;\Rightarrow\; rank(a) = rank(a').$$

*Proof*
See (Dovier et al. 2001). □

Our OBDD encoding of $G/\equiv$ is defined rank by rank as follows:

1. at each rank $i$ we determine which are the nodes reachable from all nodes at rank $i$, i.e. the list $D_i$;
2. then, we compute for each node $a$ at rank $i$ the OBDD representing $\succ_i(a)$ (two nodes $a$ and $a'$ at rank $i$ are bisimilar if and only if $\succ_i(a) = \succ_i(a')$);
3. subsequently we collect all the OBDDs obtained at rank $i$ in the list $Cod_i$ (without repetitions) and we assign to each node $a$ at rank $i$ the number $\mathbb{A}(a)$ which is the position of $\succ_i(a)$ in $Cod_i$;
4. for each node $a$ at rank $i$ its binary encoding is replaced with $\langle i, \mathbb{A}(a) \rangle$.

Hence, two nodes $a$ and $a'$ at rank $i$ are bisimilar if and only if $\mathbb{A}(a) = \mathbb{A}(a')$ and, in general, two nodes $a$ and $a'$ are bisimilar if and only if $\langle rank(a), \mathbb{A}(a)\rangle = \langle rank(a'), \mathbb{A}(a')\rangle$. For this reason we introduce the *encoding pair* (corresponding to the classical Ackermann encoding) of a node $a$ as the pair $\langle rank(a), \mathbb{A}(a)\rangle$.

*Definition 12* (*OBDD encoding - acyclic case*)
Let $G = \langle N, \succ, r\rangle$ be an acyclic graph. For each rank $i$, and each node $a$ of $G$ we define, by induction on the rank, $D_i$, $Cod_i$, $\succ_i(a)$, and $\mathbb{A}(a)$ as follows:

$$D_i = [\langle rank(b), \mathbb{A}(b)\rangle \mid \exists a(rank(a) = i \wedge a \succ b)]$$
$$\succ_i(a) = \bigvee_{a \succ b} \chi_{\{d_i(b)\}}$$
$$Cod_i = [\succ_i(a) \mid rank(a) = i]$$
$$\mathbb{A}(a) = k \quad \text{if and only if} \quad Cod_i[k] = \succ_i(a)$$

where $D_i$ and $Cod_i$ are ordered lists without repetitions, $\succ_i(a)$ is an OBDD, and $\mathbb{A}(a)$ is a natural number.

*Theorem 13*
Let $G = \langle N, \succ, r\rangle$ be an acyclic graph and $a, a' \in N$ be two nodes.

$$\langle rank(a), \mathbb{A}(a)\rangle = \langle rank(a'), \mathbb{A}(a')\rangle \quad \Leftrightarrow \quad a \equiv a'.$$

*Proof*
($\Rightarrow$) Consider the relation $B$ defined as $aBa'$ if and only if $\langle rank(a), \mathbb{A}(a)\rangle = \langle rank(a'), \mathbb{A}(a')\rangle$. We prove that $B$ is a bisimulation. If $rank(a) = rank(a') = 0$, then $nBn'$ satisfies the forward and the backward condition, since $a$ and $a'$ have no successors. If $rank(a) = rank(n') = i$ and $\mathbb{A}(a) = \mathbb{A}(a')$, then $\succ_i(a) = \succ_i(a')$, hence for each $b$ such that $a \succ b$ and $\langle rank(b), \mathbb{A}(b)\rangle$ is the $j^{th}$ element of $Dom_i$ there exists $b'$ such that $a' \succ b'$ and $\langle rank(b'), \mathbb{A}(b')\rangle$ is the $j^{th}$ element of $Dom_i$, hence we have that $bBb'$. Similarly we can prove that for each $b'$ such that $a' \succ b'$ there exists $b$ such that $a \succ b$ and $bBb'$, hence we have that $aBa'$ satisfies the forward and the backward conditions. From the fact that $B$ is a bisimulation we have that it is included in $\equiv$, i.e. the thesis.

($\Leftarrow$) By induction on the rank. If $rank(a) = 0$ and $a \equiv a'$, then from Lemma 11 we have that $rank(a') = 0$, hence $\mathbb{A}(a) = \mathbb{A}(a') = 0$. If $rank(a) = i$ and $a \equiv a'$, then from Lemma 11 we have that $rank\,(a') = i$, hence we have to prove that $\succ_i(a) = \succ_i(a')$. If $\chi_{\{d_i(b)\}}$ is a disjunct in $\succ_i(a)$, then there exists $b$ such that $a \succ b$ and $\langle rank\,(b), \mathbb{A}(b)\rangle$ is the $d_i(b)^{th}$ element of $Dom_i$. Since $a \equiv a'$, there exists $b'$ such that $a' \succ b'$ and $b \equiv b'$. By inductive hypothesis, since $rank\,(b) < i$, we obtain that $\langle rank\,(b), \mathbb{A}(b)\rangle = \langle rank\,(b'), \mathbb{A}(b')\rangle$, hence $\chi_{\{d_i(b)\}}$ is a disjunct in $\succ_i(a')$. Similarly we can prove that if $\chi_{\{d_i(b')\}}$ is a disjunct in $\succ_i(a')$, then it is also a disjunct in $\succ_i(a)$, from which $\succ_i(a) = \succ_i(a')$ and hence, $\mathbb{A}(a) = \mathbb{A}(a')$. □

*Example 14*
*Consider the graph $G$ in Figure 6.*

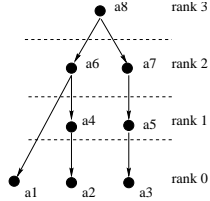

Fig. 6

*At rank $0$ we obtain:*

$$D_0 = \emptyset$$
$$\succ_0(a_1) = \succ_0(a_2) = \succ(a_3) = \bot$$
$$\mathbb{A}(a_1) = \mathbb{A}(a_2) = \mathbb{A}(a_3) = 0$$

*At rank $1$:*

$$D_1 = [\langle 0, 0\rangle] \qquad \mathcal{D}_1 = \{\langle 0, 0, 0\rangle\}$$
$$\succ_1(a_4) = \succ_1(a_5) = \neg z_1 \quad \mathbb{A}(a_4) = \mathbb{A}(a_5) = 0$$

*At rank $2$:*

$$D_2 = [\langle 0,0\rangle, \langle 1,0\rangle] \qquad \mathcal{D}_2 = \{\langle 0,0,0\rangle, \langle 1,0,1\rangle\}$$
$$\succ_2(a_6) = \neg z_1 \vee z_1,\ \succ_2(a_7) = z_1 \quad \mathbb{A}(a_6) = 0,\ \mathbb{A}(a_7) = 1$$

*At rank $3$:*

$$D_3 = [\langle 2,0\rangle, \langle 2,1\rangle] \quad \mathcal{D}_3 = \{\langle 2,0,0\rangle, \langle 2,1,1\rangle\}$$
$$\succ_3(a_8) = \neg z_1 \vee z_1 \quad \mathbb{A}(a_8) = 0$$

Let us describe the steps of our algorithm in the acyclic case.

*Algorithm 1 (Acyclic case)*
      1. **for** $a \in N$ **do compute** $rank\,(a)$; — *compute the rank*
      2. $\rho := \max\{rank\,(a) \mid a \in N\}$;
      3. **for** $i = 0, \ldots, \rho$ **do** $B_i := \{a \in N \mid rank\,(a) = i\}$;
      4. **for** $i = 0, \ldots, \rho$ **do**
            (a) $D_i := [\langle rank\,(b), \mathbb{A}(b)\rangle \mid \exists a(a \in B_i \wedge a \succ b)]$; — *determine $D_i$*
            (b) **for** $\langle rank\,(b), \mathbb{A}(b)\rangle \in D_i$ **do**
                    $d_i(b) = $ position of $\langle rank\,(b), \mathbb{A}(b)\rangle$ in $D_i$;

(c) **for** $a \in B_i$ **do**
$\succ_i(a) = \chi_{\{d_i(b) \mid a \succ b\}}$; — *compute the OBDD for a*
(d) $Cod_i = [\succ_i(a) \mid a \in B_i]$; — *collect the OBDDs at rank i*
(e) **for** $a \in B_i$ **do**
$\mathbb{A}(a) =$ position of $\succ_i(a)$ in $Cod_i$; — *compute the encoding for* a

Notice that in the algorithm $\succ_i(a)$ is computed as the OBDD of the function $\chi_{\{d_i(b) \mid a \succ b\}}$. This is equivalent to consider the OBDD of $\bigvee_{a \succ b} \chi_{\{d_i(b)\}}$. In order to compute the latter we need a procedure which from the OBDDs representing $f$ and $g$ computes the OBDD representing $f \vee g$, while the first does not require the use of such a procedure.

*Theorem 15*
Let $G = \langle N, \succ, r \rangle$ be an acyclic graph. Let $symbsize(\succ_i(a))$ be the number of nodes in the OBDD representing $\succ_i(a)$ and $symbsize(\succ) = \sum_{a \in N} symbsize(\succ_i(a))$. Algorithm 1 correctly computes the OBDD encoding of $G/\equiv$ w.r.t. Definition 12 with a worst case time $O(|N|^2)$. The worst case space complexity is $O(|N| + |\succ| + symbsize(\succ))$.

*Proof*
The correctness of the algorithm follows from Theorem 13.

Step (1.) can be performed in time $O(|N| + |\succ|)$ using a visit of the graph (see (Dovier et al. 2001)). Step (2.) can be performed in time $\rho$, which in the worst case is $|N| - 1$. Step (3.) can be performed in time $O(|N|)$. Globally all the lists $D_i$ for $i = 0, \ldots, \rho$ have a length of $|\succ|$, since each edge is used to put an element in one of the lists exactly once (i.e. when the rank of its starting node is reached). In order to avoid to add an element twice in a list $D_i$ it is sufficient to keep a global list of the $\langle rank(b), \mathbb{A}(b) \rangle$ with a flag specifying whether during the $i$th iteration the element has already been taken or not. Hence, globally all the steps (4.a) take time $O(|\succ|)$. Similarly all the steps (4.b) take time $O(|\succ|)$. The computation of $\succ_i(a)$ takes time $O(2^{log h_i})$, where $h_i = |H_i|$ (see (Clarke et al. 1999)). In the worst case all the $h_i$ are $|N|$, hence in the worst case the computation of $\succ_i(a)$ costs $O(|N|)$. So globally all the steps (4.c) cost $O(|N|^2)$. If we keep all the OBDDs in a unique table, we obtain that globally all the steps (4.d) and (4.e) cost $O(|N|)$. Hence, since $|\succ| \leq |N|^2$, the worst case time complexity is $O(|N|^2)$.

The worst case space complexity result is trivial. □

Notice that the encoding we obtain in the acyclic case is an encoding *à la* Ackermann (see (Ackermann 1937)) with two extra ingredients:

1) since we are encoding only a particular graph, and not all the possible acyclic graphs as in the general case, we keep the encoding more compact (it is exactly to this end that we use the notion of rank);
2) we use OBDDs, instead of an iterated exponential function, to compute the encoding. The use of an iterated exponential function would be very inefficient because of its fast growing. Moreover, we do not need it, since, as explained in the previous point, we use are encoding only one graph. The connections between the Ackermann encoding and our encoding are further discussed in (Piazza 2002).

## 8 Cyclic case

In order to generalize our encoding to the cyclic case, first of all we need a generalization of the notion of rank. We give here the definition of such a generalization which has been introduced in (Dovier et al. 2001).

Given a graph $G = \langle N, \succ, r \rangle$, let $G^{scc} = \langle N^{scc}, \succ^{scc}, c(r) \rangle$ be the graph of the *strongly connected components*, where $c(r)$ is the strongly connected component of $r$. Given a node $a \in N$, we refer to the node of $G^{scc}$ associated to the strongly connected component of $a$ as $c(a)$. Observe that $G^{scc}$ is acyclic and if $G$ is acyclic then $G^{scc}$ is $G$ itself.

We need to distinguish between the *well-founded* part and the *non-well-founded* part of a graph $G$.

*Definition 16 (Well-founded part)*
Let $G = \langle N, \succ, r \rangle$ and $a \in N$. $G(a) = \langle N(a), \succ_{|N(a)}, a \rangle$ is the subgraph of $G$ of the nodes reachable from $a$. $WF(G)$, the well-founded part of $G$, is $WF(G) = \{a \in N : G(a) \text{ is acyclic}\}$.

The following is an extension of the previous notion of rank (cf. Definition 10) suitable for dealing with the cyclic case.

*Definition 17 (Rank - general case)*
Let $G = \langle N, \succ, r \rangle$. The *rank* of a node $a$ of $G$ is defined as:

$$rank(a) = \begin{cases} 0 & \text{if } a \text{ is a leaf in } G \\ -1 & \text{if } c(a) \text{ is a leaf in } G^{scc} \text{ and } a \text{ is not a leaf in } G \\ \max(\{1 + rank(b) : c(a) \succ^{scc} c(b), b \in WF(G)\} \cup \\ \quad \{rank(b) : c(a) \succ^{scc} c(b), b \notin WF(G)\}) & \text{otherwise} \end{cases}$$

Since $G^{scc}$ is always acyclic, the definition is correctly given. If $G$ is acyclic then $G = G^{scc}$ and the above definition reduces to the one given in the acyclic case (Definition 10). Notice that if the graph is strongly connected all the nodes have rank $-1$. This is never the case when our graph $G$ is obtained from a labeled graph $G'$.

*Lemma 18*
Let $G$ be a graph and $a, a' \in N$ be two nodes of $G$.

$$a \equiv a' \quad \Rightarrow \quad rank(a) = rank(a').$$

*Proof*
See (Dovier et al. 2001). □

Notice that the following also holds:

$$a \succ b \quad \Rightarrow \quad rank(b) \leq rank(a).$$

In particular, as pointed out in (Dovier et al. 2001), in order to determine if two nodes at rank $i$ are bisimilar it is sufficient to know the bisimulation $\equiv$ on the nodes of rank less than $i$ and the edges among the nodes at rank $i$.

The main idea behind the extension of our way to compute the OBDD encoding in the cyclic case is that of assigning to each node *two* encodings. This means that we treat $a \succ b$ in an *asymmetric* way with respect to $\succ$: a "trick" first proposed by Fraenkel and Mostowski in their *permutation of the universe* technique (see (Jech 1978)) to model cyclic membership relations. In particular, the two encodings are $\succ_i(a)$ and $d(a)$ and we compute $\succ_i(a)$ (lhs of $a \succ b$) using $d(b)$ (rhs of $a \succ b$).

Let us assume that for each $j < i$ we have correctly encoded all the nodes of rank $j$, i.e. we have assigned to each node $a$ of rank $j$ a number $\mathbb{A}(a)$ in such a way that

$$a \equiv a' \iff \langle rank\,(a), \mathbb{A}(a) \rangle = \langle rank\,(a'), \mathbb{A}(a') \rangle.$$

We want to extend the encoding to the nodes at rank $i$. First we compute $D'_i$, all the $\succ'_i(a)$'s, $Cod'_i$, and all the $\mathbb{A}'(a)$'s as in the acyclic case, but without considering the nodes at rank $i$. This means that $\succ'_i(a)$ and $\mathbb{A}'(a)$ are a first approximation of $\succ_i(a)$ and $\mathbb{A}(a)$ in which we consider only the edges reaching a node $b$ whose rank is less than $i$. In particular:

$$D'_i = [\langle rank\,(b), \mathbb{A}(b) \rangle \mid \exists a (rank\,(a) = i \wedge a \succ b \wedge rank\,(b) < i)],$$

$$\succ'_i(a)(\bar{z}) = \bigvee_{a \succ b \wedge b \in D'_i} \chi_{\{d_i(b)\}}(\bar{z}),$$

$$Cod'_i = [\succ'_i(a) \mid rank\,(a) = i], \text{ and}$$

$$\mathbb{A}'(a) = k \quad \text{if and only if} \quad Cod'_i[k] = \succ'_i(a).$$

where $D'_i$ and $Cod'_i$ are lists without repetitions.

For each node $a$ at rank $i$ consider a variable $\mathsf{W}_{\mathsf{a}}^{\succ i}$ (which ranges over boolean functions) and a variable $\mathsf{d}_{\mathsf{a}}$ (which ranges over natural numbers). If there are $r_i$ nodes at rank $i$, then we impose

$$\mathsf{d}_{\mathsf{a}} \in \{|D'_i| + 1, \ldots, |D'_i| + r_i\}.$$

We consider all the boolean equations

$$\mathsf{W}_{\mathsf{a}}^{\succ i}(\bar{z}) = \left( \bigvee_{a \succ b \wedge rank\,(b) = i} \chi_{\{\mathsf{d}_{\mathsf{b}}\}}(\bar{z}) \right) \vee \succ'_i(a)(\bar{z}) \tag{1}$$

Notice that the second disjunct in the definition of $\mathsf{W}_{\mathsf{a}}^{\succ i}$ is a boolean function, while we cannot explicitly write the first part until we know the values of the $\mathsf{d}_{\mathsf{a}}$'s. Notice also that it is possible to replace the boolean equation with a numeric equation using the following definition

$$\mathsf{W}_{\mathsf{a}}^{\succ i} = \sum_{j=|D'_i|+1}^{|D'_i|+r_i} 2^j * min\left( \sum_{a \succ b \wedge rank\,(b) = i} 1 - \frac{|\mathsf{d}_{\mathsf{b}} - j|}{max(|\mathsf{d}_{\mathsf{b}} - j|, 1)}, 1 \right) + 2^{d'(a)},$$

with $d'(a) = \mathbb{A}'(a) + |D'_i| + r_i + 1$. This means that if $S = \{\gg_i(a), d_i(a) \,|\, rank\,(a) = i\}$ is a solution of this system of boolean equations, then $\gg_i(a)$ is the OBDD associated to $a$ and $d(a)$ is the position of $a$ in $D_i$. Moreover, we consider all the boolean equations of the form

$$(\mathtt{W}_\mathtt{a}^{>i} \leftrightarrow \mathtt{W}_{\mathtt{a}'}^{>i}) \quad \leftrightarrow \quad \mathtt{d}_\mathtt{a} = \mathtt{d}_{\mathtt{a}'} \tag{2}$$

which can be expressed as numeric equations as

$$\frac{|\mathtt{d}_\mathtt{a} - \mathtt{d}_{\mathtt{a}'}|}{max(|\mathtt{d}_\mathtt{a} - \mathtt{d}_{\mathtt{a}'}|, 1)} \;=\; \frac{|\mathtt{W}_\mathtt{a}^{>i} - \mathtt{W}_{\mathtt{a}'}^{>i}|}{max(|\mathtt{W}_\mathtt{a}^{>i} - \mathtt{W}_{\mathtt{a}'}^{>i}|, 1)}.$$

Let $Sys_i$ be the system containing all the equations (1) and (2). We put as objective function to be minimized

$$max\{\mathtt{d}_\mathtt{a} \,|\, rank\,(a) = i\},$$

i.e. we want to maximize the number of equalities between the $\mathtt{d}_\mathtt{a}$'s. Let $S = \{\gg_i(a), d_i(a) \,|\, rank\,(a) = i\}$ be a solution of the system $Sys_i$ which minimizes the objective function. We assign to each node $a$ of rank $i$ the code

$$\mathbb{A}(a) \;=\; d_i(a) - (|D'_i| + 1).$$

Notice that we always obtain that all the nodes $a$ at rank $-1$ have $\mathbb{A}(a) = 0$. This is correct, since we are working on unlabeled graphs.

In order to show that our technique is correct we begin showing that the system has always at least one solution. We prove this by proving that from the maximal bisimulation we are able to describe a solution.

*Lemma 19*
Let $G$ be a graph, and let $\equiv$ be the maximum bisimulation over $G$. Consider an ordering $ord$ (starting from 0) of the equivalence classes of $G/\equiv$ at rank $i$ and for each node $a$ at rank $i$ define

$$\begin{aligned} d_i(a) &= ord([a]) + |D'_i| + 1 \\ \gg_i(a) &= \left(\bigvee_{a \gg b \wedge rank\,(b) = i} \chi_{\{d(b)\}}\right) \vee \gg'_i(a). \end{aligned}$$

The set $S = \{\gg_i(a), d_i(a) \,|\, rank\,(a) = i\}$ is a solution of $Sys_i$.

*Proof*
We recall that we are assuming that for all the nodes at rank less than $i$ we have that

$$a \equiv a' \;\Leftrightarrow\; \langle rank\,(a), \mathbb{A}(a)\rangle = \langle rank\,(a'), \mathbb{A}(a')\rangle.$$

By induction on the rank.

Let $i = -1$. Since $\equiv$ is the maximum bisimulation all the nodes at rank $-1$ are bisimilar (see (Aczel. 1988; Piazza 2002)), hence there is only one equivalence class and all the $d(a)$ are equal to 1. We have $\gg'_{-1}(a) = \bot$ for all the nodes at rank $-1$. All the equations of the $\mathtt{W}_\mathtt{a}^{>-1}$'s are satisfied, since we use them to define the $\gg_{-1}(a)$'s in terms of the $d_{-1}(a)$'s. We have to prove that $\gg_{-1}(a) = \gg_{-1}(a') \leftrightarrow$ $d_{-1}(a) = d_{-1}(a')$. So what we have to prove is that all the $\succ_{-1}(a)$'s are equal. If $a$ is a node at rank $-1$, then there exists at least a node $b$ such that $a \succ b$ and $rank(b) = -1$. Hence, if $rank(a) = -1$ we obtain $\succ_{-1}(a) = \chi_{\{1\}}$, i.e. all the $\succ_{-1}(a)$'s are equal.

Let $i = 0$. Since $\equiv$ is the maximum bisimulation all the nodes at rank 0 are bisimilar (see (Aczel. 1988; Piazza 2002)), hence there is only one equivalence class and all the $d_0(a)$ are equal to 1. All the equations of the $\mathtt{W}_\mathtt{a}^{\succ_0}$'s are satisfied, since we use them to define the $\succ_0(a)$'s in terms of the $d_0(a)$'s. We have to prove that $\succ_0(a) = \succ_0(a') \leftrightarrow d_0(a) = d_0(a')$. Hence, what we have to prove is that all the $\succ_0(a)$'s are equal. If $a$ is a node at rank 0, then it has no outgoing edges, therefore we obtain $\succ_0(a) = \bot$, i.e. all the $\succ_0(a)$'s are equal.

Let $i = k+1$. All the equations of the $\mathtt{W}_\mathtt{a}^{\succ_i}$'s are satisfied, since we use them to define the $\succ_i(a)$'s in terms of the $d_i(a)$'s. We have to prove that $\succ_i(a) = \succ_i(a') \leftrightarrow d_i(a) = d_i(a')$. If $\succ_i(a) = \succ_i(a')$, then we have to prove that $d_i(a) = d_i(a')$, i.e. we have to prove that $a \equiv a'$. If $a \succ b$ and $rank(b) < i$, then in $\succ_i(a)$ there is a disjunct of the form $\chi_{\{d_i(b)\}}$ with $d_i(b) \leq |D_i'|$. The same disjunct is in $\succ_i(a')$, and hence we obtain that there exists $b'$ such that $b \equiv b'$ and $a' \succ b'$. If $a \succ b$ and $rank(b) = i$, then in $\succ_i(a)$ there is a disjunct of the form $\chi_{\{d_i(b)\}}$ with $d_i(b) > |D_i'|$, hence the same disjunct must be in $\succ_i(a')$, i.e. (from the definition of $d_i(b)$) there exists $b'$ such that $a' \succ b'$ and $b' \equiv b$. Similarly for the backward condition. Hence we have $a \equiv a'$, i.e. $d_i(a) = d_i(a')$. If $d_i(a) = d_i(a')$, then $a \equiv a'$. We have to prove that $\succ_i(a) = \succ_i(a')$. If in $\succ_i(a)$ there is a component of the form $\chi_{\{d_i(b)\}}$ with $d_i(b) \leq |D_i'|$, then $a \succ b$ and $rank(b) < i$, hence there exists $b'$ such that $a' \succ b'$ and $b \equiv b'$, i.e. $d_i(b) = d_i(b')$ from which we obtain that in $\succ_i(a')$ there is a component of the form $\chi_{\{d_i(b)\}}$. If in $\succ_i(a)$ there is a component of the form $\chi_{\{d_i(b)\}}$, with $d_i(b) > |D_i|$, then since $a \equiv a'$, there must be $b'$ such that $a' \succ b'$, and $b \equiv b'$, from which we obtain that in $\succ_i(a')$ there is a component of the form $\chi_{\{d_i(b)\}}$. Similarly, it is possible to prove that all the components that are in $\succ_i(a')$ are also in $\succ_i(a)$, from which we obtain that $\succ_i(a) = \succ_i(a')$. □

On the ground of the above result we can show that with the proposed encoding we obtain a (unique and symbolic) representation of $G/\equiv$.

*Theorem 20*
Let $G = \langle N, \succ, r \rangle$ be a graph and $a, a' \in N$ be two nodes.

$$\langle rank(a), \mathbb{A}(a) \rangle = \langle rank(a'), \mathbb{A}(a') \rangle \quad \Leftrightarrow \quad a \equiv a'.$$

*Proof*
($\Rightarrow$) Let $B$ be defined as $aBa'$ if and only if $\langle rank(a), \mathbb{A}(a) \rangle = \langle rank(a'), \mathbb{A}(a') \rangle$. We prove that $B$ is a bisimulation. If $rank(a) = -1$ and $aBa'$, then we have nothing to prove since all the equivalence relations between nodes at rank $-1$ are bisimulations. Also the case in which $rank(a) = 0$ is trivial. Let $rank(a) = i$ and $aBa'$. If $a \succ b$ and $rank(b) < i$, then in $\succ_i(a)$ there is a disjunct of the form $\chi_{\{d_i(b)\}}$ with $d_i(b) \leq |D_i'|$. The same component occurs in $\succ_i(a')$, hence there must exists $b'$ such that $a' \succ b'$, $rank(b') < i$ and $\langle rank(b), \mathbb{A}(b) \rangle = \langle rank(b'), \mathbb{A}(b') \rangle$, from which we obtain that $b'Bb$. If $a \gg b$ and $rank\,(b) = i$, then in $\gg_i(a)$ there is a component of the form $\chi_{\{d_i(b)\}}$ with $d_i(b) > |D_i'|$. Since $\gg_i(a') = \gg_i(a)$ we obtain that in $\gg_i(a')$ there is a component of the form $\chi_{\{d_i(b)\}}$. This implies that there exists $b'$ such that $a' \gg b'$, $rank\,(b') = i$ and $d_i(b) = d_i(b')$ in the solution of the system. From the fact that $d_i(b) = d_i(b')$ we obtain $\gg_i(b) = \gg_i(b')$, hence $bBb'$. The backward condition is similar.

($\Leftarrow$) From Lemma 19 we have that the maximal bisimulation gives us a solution. From each solution we are able to define a bisimulation $B$ such that

$$max\{d_i(a) \mid rank\,(a) = i\} + 1$$

is the number of classes at rank $i$ (see ($\Rightarrow$)). Hence, it must be that the solution obtained from $\equiv$ minimizes $max\{d_i(a) \mid rank\,(a) = i\}$ at each rank. $\square$

Notice that in the numeric system we use the exponential function $2^{d_b}$ in order to compute $\gg_i(a)$. We could have used any other function $f$ such that:

- $f([b, b|R]) = f([b|R])$ ($f$ does not depend on the number of repetitions);
- $f([b_1, b_2, \ldots, b_h]) = f([b_{i1}, b_{i2}, \ldots, b_{ih}])$, where on the r.h.s. we have a permutation of the l.h.s. ($f$ does not depend on the ordering);
- $f(L1) \neq f(L2)$, if in $L1$ there is an element which is not in $L2$ or viceversa.

*Example 21*
*Let us assume that at rank $i$ we have the sub-graph presented in Figure 7 and that $\gg_i'(a) = \gg_i'(c) = f(\bar{z})$ and $\gg_i'(b) = \gg_i'(d) = g(\bar{z})$. In this case a minimal solution*

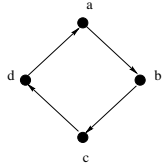

Fig. 7

*of the system of boolean equation is given by*

$$d_i(a) = d_i(c) = |D_i'| + 1, \quad d_i(b) = d_i(d) = |D_i'| + 2,$$

*and hence we can assign*

$$\mathbb{A}(a) = \mathbb{A}(c) = 0, \mathbb{A}(b) = \mathbb{A}(d) = 1.$$

Let us describe the steps of our algorithm in the cyclic case.

*Algorithm 2 (Cyclic case)*
    1. **for** $a \in N$ **do compute** $rank\,(a)$; — *compute the rank*
    2. $\rho := \max\{rank\,(a) \mid a \in N\}$;
    3. $P := \{B_i \mid i = 0, \ldots, \rho\}$;
    4. **for** $i = 0, \ldots, \rho$ **do**
        (a) $D_i' := [\langle rank\,(b), \mathbb{A}(b)\rangle \mid \exists a(a \in B_i \wedge a \gg b \wedge rank\,(b) < i)]$;

- (b) **for** $\langle rank\,(b), \mathbb{A}(b) \rangle \in D'_i$ **do**
  $d_i(b) = $ position of $\langle rank\,(b), \mathbb{A}(b)\rangle$ in $D'_i$;
- (c) **for** $a \in B_i$ **do**
  $\succ'_i(a) = \chi_{\{d_i(b)\,|\,a \succ b \wedge rank\,(b) < i\}}$;
- (d) $Cod'_i = [\succ'_i(a)\,|\,a \in B_i]$;
- (e) **for** $a \in B_i$ **do**
  $\mathbb{A}'(a) = $ position of $\succ'_i(a)$ in $Cod'_i$;

5. $Sys_i = \emptyset$;
6. **for** $a \in B_i$ **do**
   $Sys_i = Sys_i \wedge \mathtt{W}_\mathtt{a}^{\succ_i} = \chi_{\{\mathtt{d}_\mathtt{b}\,|\,a \succ b \wedge b \in B_i\}} \vee \succ'_i(a)$;
7. **for** $a, a' \in B_i$ **do**
   $Sys_i = Sys_i \wedge (\mathtt{W}_\mathtt{a}^{\succ_i} \leftrightarrow \mathtt{W}_{\mathtt{a}'}^{\succ_i}) \leftrightarrow \mathtt{d}_\mathtt{a} = \mathtt{d}_{\mathtt{a}'}$;
8. $S = $ a solution of $Sys_i$ minimizing $max\{\mathtt{d}_\mathtt{a}\,|\,a \in B_i\} = \{\succ_i(a), d_i(a)\,|\,a \in B_i\}$;
9. **for** $a \in B_i$ **do**
   $\mathbb{A}(a) = d_i(a) - (|D'_i| + 1)$;

*Theorem 22*
Let $G = \langle N, \succ, r\rangle$ be a graph. Let $symbsize(\succ_i(a))$ be the number of nodes in the OBDD representing $\succ_i(a)$ and $symbsize(\succ) = \sum_{a \in N} symbsize(\succ_i(a))$. Algorithm 2 correctly computes the OBDD encoding of $G/\equiv$ with a worst case time $O(|N|^2 + |\succ|\log|N|)$. The worst case space complexity is $O(|N| + |\succ| + symbsize(\succ))$.

*Proof*
The correctness of the algorithm follows from Theorem 20.

Steps (1.)–(4.) have a worst case time complexity $O(|N|^2$. As far as steps (5.)–(8.) are concerned, notice that we are only interested in the solution of $Sys_i$. As it follows from Lemma 19 the solution $S$ can be determined using the maximum bisimulation. This can be found using Paige-Tarjan algorithm (Paige and Tarjan 1987) in time $O(|\succ|\log|N|)$. From the maximum bisimulation we can build $\succ_i(a)$ in time $O(|N|)$. Hence, globally all the steps (5.)–(8.) take time $O(|\succ|\log|N| + |N|^2)$. It is plain that all the steps (9.) take time $O(|N|)$.

The worst case space complexity result is trivial. □

Notice that in order to obtain the solutions of all the $Sys_i$'s in time $O(|\succ|\log|N| + |N|^2)$ we exploit Paige-Tarjan (Paige and Tarjan 1987) algorithm. This is possible since $Sys_i$ is nothing but a boolean encoding of the bisimulation problem for the nodes at rank $i$. Having an algorithm which computes the maximum bisimulation in time $T$ we can find the solutions of all the $Sys_i$'s in time $T + O(|N|^2)$, where the $O(|N|^2)$ time complexity is due to the construction of all the OBDDs $\succ_i(a)$.

## 9 Final Considerations

Given a graph $G$ our method allows to compute a symbolic representation of $G/\equiv$. Hence, the smaller is $G/\equiv$ with respect to $G$ the smaller is our representation with respect to the one obtained by applying the standard method (briefly described in Section 4). To show this fact let us consider some extreme cases. Let $G$ be a graph constituted by $n$ self-loops. Since the $n$ nodes are all bisimilar $G/\equiv$ has only one node with one self-loop and our representation of $G/\equiv$ requires only one OBDD with 2 internal nodes. The size of the standard OBDD representation of $G$ grows with $n$. For instance, with $n = 4$ the OBDD has 9 internal nodes, while with $n = 8$ it has 19 internal nodes. Similarly, if we consider a graph $G$ of $n$ nodes constituting a loop, we get that, since $G/\equiv$ has only one node and one self-loop, our representation requires only one OBDD with 2 internal nodes. Again the size of the standard representation grows with $n$. For instance, if $n = 4$, then the OBDD has 8 nodes, while with $n = 8$ the OBDD has 21 nodes. We obtain similar results also considering graphs constituted by chains of nodes.

Other cases in which our representation uses less space than the standard one can be easily obtained by considering graphs in which the same structure is repeated at different ranks. In fact, in these cases we can reuse the same OBDDs at different ranks.

By combining the two above observations, we can build classes of graphs in which the same structure is repeated at different ranks of the bisimulation quotient. Also in these cases we can reuse the same OBDDs at different ranks.

Recently a joint CWI/INRIA project has been started to establish an "official" benchmark suite for large transition systems. A preliminary version of the benchmark suite, called VLTS (Very Large Transition Systems) Benchmark Suite, is available at `http://www.inrialpes.fr/vasy/cadp/resources/benchmark_bcg.html`. The benchmarks have been obtained from various case studies modelling communication protocols and concurrent systems. Many of these case studies correspond to real life, industrial systems. Two peculiarities of the benchmarks (see also their graphical representations in the web page) suggest that we can strongly exploit the OBBDs reuse typical of our encoding. Such characteristics correspond to: 1) the presence of deadlocks and; 2) a low number of outgoing edges per nodes. From our point of view the presence of deadlocks ensures that there are nodes at different ranks, since all the deadlocks are nodes of rank 0, all the nodes which can reach a deadlock in one step are at rank 1, and so on. Moreover, a low number of outgoing edges from a node increases the probability of using the same OBDDs to represent nodes at different ranks (see Section 5). Experimentation is in progress.

## 10  Conclusions

In this paper we first proposed a way to represent graphs via OBDDs *provided* they are reduced with respect to the maximum bisimulation $\equiv$. Our technique is based on the observation that, in a graph quotiented by $\equiv$, each node can be uniquely characterized by the set of its successors. Moreover, the representation we propose can be computed on a graph *layered* with respect to an equivalence relation. This allows, in general, sharing among the employed OBDDs, thereby guaranteeing savings in space.

The second part of the paper shows how, when the layering is based on a suitable notion of rank, the representation can be used as an Ackermann encoding. Such an encoding allows a computation of the quotient structure *together with* its OBDDs representation. This second part begins with the treatment of acyclic graphs and then provides the representation in the general case as a solution of a system of equations. The encoding we propose, by exploiting the (*a priori*) knowledge on the size of the graph (set), greatly reduces the size of the involved numbers with respect to the standard Ackermann encoding.

It should be noted that the Ackermann encoding is, in fact, very similar to the OBDD representation of a graph: both techniques are designed to provide a compact representation of a graph and are based on a binary representation. The main difference is the inability of the OBDD representation to fully exploit the space savings induced by bisimulation, as the only sharing automatically introduced by the OBDD representation allows avoiding repetition of nodes having equal *immediate* successors. On the other hand, however, the Ackermann encoding was a map designed (for entirely different purposes) to embed the *entire* universe of well-founded hereditarily finite sets, hence the numbers involved tend to grow very fast and the cyclic sets could not be mapped.

In this paper we have tried to retain the good aspects of both techniques introducing a method integrating Ackermann idea of (uniquely) encoding sets by rank and representing the encoding by OBDD's instead of natural numbers.

A similar approach for the computation of the quotient w.r.t. the maximum simulation has been considered in (Gentilini et al. 2001; Piazza 2002). The main difference in the case of simulation is that while considering the nodes at rank $i$ it is also necessary to update the encodings of the nodes at rank less than $i$ in an *incremental* way. The encoding presented in (Gentilini et al. 2001; Piazza 2002) differs from the one presented here because it encodes for each node $a$ the set of nodes which are simulated by $a$. We are currently investigating the computation of the simulation quotient exploiting the encoding presented here.

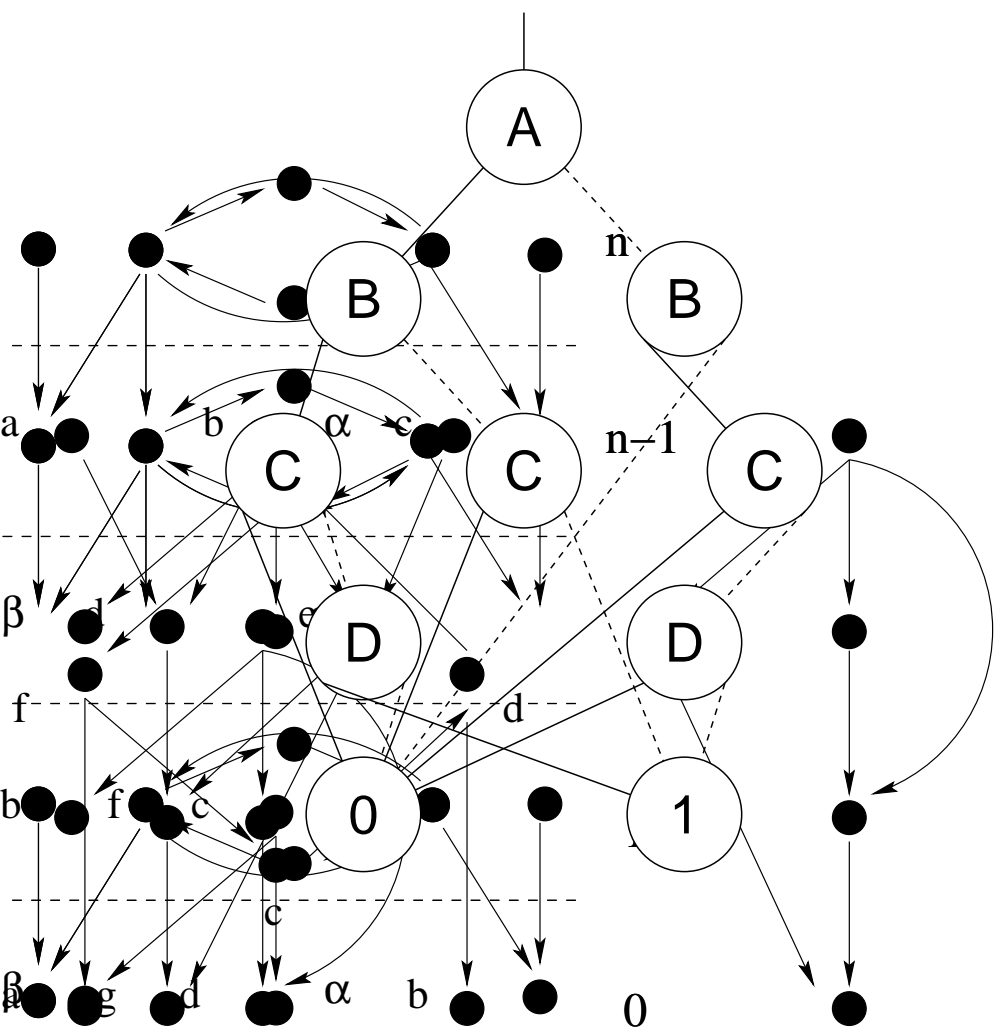